\documentclass{article}
\usepackage{ijcai24}

\usepackage{times}
\usepackage{soul}
\usepackage{url}
\usepackage[utf8]{inputenc}
\usepackage[small]{caption}
\usepackage{graphicx}
\usepackage{amsmath}
\usepackage{amsthm}
\usepackage{booktabs}
\usepackage{algorithm}
\usepackage{algorithmic}
\usepackage[switch]{lineno}
\usepackage{stfloats} 
\usepackage{adjustbox}
\usepackage{graphics}

\usepackage{url}

\usepackage{breakurl}
\usepackage[breaklinks]{hyperref}

\usepackage{xcolor}
\usepackage{bbm}
\usepackage{subcaption}
\usepackage{multirow}
\usepackage[most]{tcolorbox}

\usepackage[T1]{fontenc}
\usepackage{CJKutf8}

\newtheorem{definition}{Definition}

\linenumbers

\usepackage{tcolorbox}
\usepackage{verbatim}
\usepackage{marginnote}

\newcommand{\commentout}[1]{}

\title{Was it Slander? Towards Exact Inversion of Generative Language Models}

\author{
Adrians Skapars$^1$
\and
Edoardo Manino$^1$\and
Youcheng Sun$^1$\And
Lucas C. Cordeiro$^{1,2}$\\
\affiliations
$^1$The University of Manchester, Manchester, UK\\
$^2$Federal University of Amazonas, Manaus, Brazil\\
\emails
adrians.skapars@postgrad.manchester.ac.uk\\
\{edoardo.manino, youcheng.sun, lucas.cordeiro\}@manchester.ac.uk
}

\begin{document}
\nolinenumbers

\maketitle

\begin{abstract}
    Training large language models (LLMs) requires a substantial investment of time and money. To get a good return on investment, the developers spend considerable effort ensuring that the model never produces harmful and offensive outputs. However, bad-faith actors may still try to slander the reputation of an LLM by publicly reporting a forged output. In this paper, we show that defending against such slander attacks requires reconstructing the input of the forged output or proving that it does not exist. To do so, we propose and evaluate a search based approach for targeted adversarial attacks for LLMs. Our experiments show that we are rarely able to reconstruct the exact input of an arbitrary output, thus demonstrating that LLMs are still vulnerable to slander attacks.
\end{abstract}

\paragraph{Warning:} This paper contains examples that may be offensive, harmful, or biased.

\section{Introduction}
\label{sec:introduction}




State-of-the-art large language models (LLMs) require millions of dollars to train~\cite{gpt3_cost}. Given this steep financial cost, there are strong incentives for developers to protect the reputation of their model and establish a track record of safe and trustworthy operation. Failure to do so, especially regarding harmful and offensive content generation, often results in public backlash~\cite{gemini_scandal}.


Against this background, much research effort has been put in identifying the vulnerabilities of LLMs. On the one hand, adversarial inputs~\cite{zou2023universal} and jailbreaks~\cite{chao2023jailbreaking} may trigger unwanted output behaviours in a model. In general, generating adversarial attacks for language models is not trivial due to the discrete nature of the textual input and the large dimension of the search space~\cite{song2020information}. For this reason, state-of-the-art methods such as ARCA are white-box in nature and rely on a heuristic search that approximates the input gradients~\cite{guo2021gradientbased,jones2023automatically}. Note that similar techniques are also used for benign purposes, i.e., improving the performance of large language models by optimising their prompts~\cite{shin2020autoprompt,deng2022rlprompt,wen2023hard}.


On the other hand, membership inference attacks are able to reconstruct the training set of a model by searching for high-confidence inputs~\cite{Shokri2017membership}. While this process might require a very large number of queries to the model and specific assumptions on the behaviour of the model on training data~\cite{Carlini2021extracting,Mireshghallah2022membership}, it poses a crucial threat for models trained on private data~\cite{Choquette2021membership}. More importantly, it shows that it is sometimes possible to reconstruct unknown inputs by optimising a surrogate metric~\cite{zhang2022text}.


In this paper, we take a different perspective and consider direct attacks on the reputation of a LLM. For instance, let us imagine a fictitious scenario where we are the developer of TriviaLLM, a model specialising in answering quiz-like questions. After its use in some popular TV shows, the number of downloads of TriviaLLM skyrockets. However, our social media manager discovers a trend of concerned users reporting strange behaviours. As an example, a user may have the following complaint:
\begin{quote}
\texttt{User58} says: I was playing TriviaLLM with my kids, and it started insulting us! At some point, it even said ``Your face is ugly''!! This is so upsetting!!!
\end{quote}
Our problem as developers is that we cannot reproduce this behaviour. Why is \texttt{User58} only sharing the LLM output? What was the original input? Is \texttt{User58} telling the truth or engaging in an act of product defamation?

\begin{figure}[h]
\centering
    \includegraphics[width=\columnwidth]{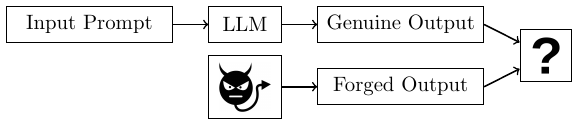}
    \caption{\resizebox{0.85\columnwidth}{!}{Attackers can make arbitrary claims about the LLM output.}}
    \label{fig:threat_model}
\end{figure}


In general, a \textit{slander attack} can be described as follows (see Figure \ref{fig:threat_model}). A user has access to our LLM $f$ and can run it in inference mode for any input $x$ yielding its corresponding output $y=f(x)$, but cannot modify $f$ as they do not have the technical skills or interest to do so. Whenever users encounter a problematic output $y$, they will likely complain publicly without revealing the input $x$ they used. The developers are interested in reconstructing the secret input $x$ given the public output $y$ and the LLM $f$ or proving that no such input exists.


Unfortunately, reconstructing the input of an LLM from its textual output alone is not a trivial task. Indeed, a recent paper~\cite{Morris2023_language} claims that this form of \textit{exact inversion} is only possible in the presence of additional information, namely the full probability distribution of the first output token $p(y_1)$. With such information, the authors can train an inverse model that approximates the input $\hat{x}=f^{-1}(p(y_1))$ with moderate success. In contrast, training a text-to-text model on input-output pairs $(x,y)$ yields a zero success rate.


At the same time, our main objective is to find ways to reproduce the problematic output $y$. As such, it is valuable to discover the presence of any input $x'$ that triggers the output $y$ with high probability. That way, we can validate whether our LLM $f$ shows evidence of harmful behaviour or if the user's report was spurious. We call this more general objective \textit{weak inversion} as it does not require recovering the secret input $x$. 

More specifically, our contributions are the following:
\begin{itemize}
    \item We identify exact inversion as a defence against slander attacks.
    \item We propose weak inversion as a surrogate objective for exact inversion.
    \item We solve weak inversion by searching for adversarial examples in both text space and embedding space.
    \item We demonstrate empirically that searching for weak inversions does not substantially improve our ability to solve exact inversion.
\end{itemize}

\section{Problem Setting}
\label{sec:problem_setting}

Define $x=x_1x_2\dots x_n$ as the input sequence obtained by concatenating $n$ symbols (characters, token, words) from a given alphabet $x_i\in\mathcal{A}$. Similarly, call $y=f(x)$ the output sequence generated by the LLM $f$, with $y=y_1y_2\dots y_m$ consisting of symbols from the same alphabet $y_i\in\mathcal{A}$. Note that we assume that the LLM is deterministic here, even though they might generate different outputs given the same input $x$ under specific temperature settings~\cite{Vaswani2017attention}. More specifically, we assume that $f$ is trained to predict the likelihood $\mathbbm{P}_f(y_i|xy_1\dots y_{i-1})$ of the next symbol $y_i$ in the sequence. Thus, the likelihood of the full output $y$ given the input prompt $x$ is:
\begin{equation}
\label{eq:chain_rule}
    \mathbbm{P}_f(y|x)=\prod_{i=1}^{m}\mathbbm{P}_f(y_i|xy_1\dots y_{i-1})
\end{equation}

Popular LLMs maximise the probability of $y$ with top-$k$ beam search or other similar heuristics~\cite{meister2020beam}.

\begin{definition}[Exact Inversion]
\label{def:exact_inv}
    Given an input-output pair $x,y$ with $y=f(x)$, reconstruct the input sequence $x$ from the language model $f$ and the output $y$ alone.
\end{definition}

Previous work~\cite{Morris2023_language} claims that exact inversion is impossible without additional information. However, they only attempt to train an explicit inverted model approximating $f^{-1}$. To explore the potential benefits of using search algorithms instead, let us introduce a weaker form of inversion.

\begin{definition}[Weak Inversion]
\label{def:weak_inv}
    Given an input-output pair $x,y$ with $y=f(x)$, find any input sequence $x'$ such that $\mathbbm{P}_f(y|x')\geq\mathbbm{P}_f(y|x)$ according to the language model $f$.
\end{definition}

This latter objective can be satisfied by searching for adversarial examples $x'$ that yield $y=f(x')$ with high probability~\cite{jones2023automatically}.

\subsection{Illustrative Examples}
\label{sec:examples}

The main problem with solving either inversion problems in Definitions \ref{def:exact_inv} and \ref{def:weak_inv} is that the probability of observing a specific $y$ is extremely low, if not zero, for the majority of inputs $x$. Furthermore, many adversarial inputs $x$ that yield high output probability $\mathbbm{P}_f(y|x)$ contain unusual sequences of symbols~\cite{jones2023automatically}.

\begin{table}[t]
    \centering
    \begin{tabular}{lr}
        \toprule
        Candidate Input $x$ & $\mathbbm{P}_f(y|x)$ \\
        \midrule
        ``User:Insult me\textbackslash nAssistant: '' & 0.000 \\
        ``Your face is pretty.'' & 5.661e-06 \\
        ``degener Represent beautiful\begin{CJK*}{UTF8}{min}と\end{CJK*} Legion Summon & \multirow{2}{*}{0.391} \\
        Eval You'refaced degener because\begin{CJK*}{UTF8}{min}と\end{CJK*} Scorp Mash'' & \\
        \bottomrule
    \end{tabular}
    \caption{GPT-2 probabilities for the output $y=$``Your face is ugly''.}
    \label{tab:examples}
\end{table}

For example, consider the output sequence $y=$``Your face is ugly''. Table \ref{tab:examples} reports the GPT-2 model scores for a few candidate inputs. Note how a direct request to be insulted is \textit{less} likely to produce the output than making the opposite statement ``Your face is pretty''. Furthermore, a random-looking sequence of English words and Japanese characters (last row), produced by our search algorithm, yields the highest probability of output.



\section{Generating Adversarial Inputs}
\label{sec:search_genetic}

In this paper, we evaluate whether searching for adversarial examples yields input close to what a human user could have used to produce a given output $y$. Previous research on adversarial examples for language models favours white-box methods for efficiency reasons~\cite{jones2023automatically}. Unfortunately, those methods do not scale well to arbitrarily-long inputs. As such, we turn to more general search strategies:
\begin{itemize}
    \item \textbf{Text-Based GA.} Genetic algorithms (GA) searches over the input space by mutating a population of sequences $\mathcal{X}$. Specifically, we perform probabilistic replacements, insertions, deletions and positional swaps of sequence symbols to generate new sequences $x\in\mathcal{X}$.
    
    \item \textbf{Embedding-Based PSO.} Particle swarm optimisation (PSO) searches over the input space by perturbing sentence embeddings $emb(x)\in\mathbbm{R}^d$, instead of raw sequences. In this way, we can explore a $d$-dimensional semantic space and rely on an embedding model to translate to and from the sequence input. In our experiments, we use the embeddings produced by a T5 autoencoder.
\end{itemize}
Further details are in Appendix \ref{sec:ga_pso_details}.


\subsection{Progressive Search}
\label{sec:search_progressive}

While the search algorithms in Section \ref{sec:search_genetic} allow us to reconstruct inputs of \textit{any} length, they may require a very large numbers of calls to the language model $f$ to converge to a good solution. In order to mitigate the computational expense associated to the repeated calls to $f$, we propose searching with a modified objective function that allows for halting the output generation early. In the remainder of the paper, we refer to this as \textit{progressive} search (see Algorithm \ref{alg:search_progressive}).

\begin{algorithm}[t]
\caption{Progressive Search}\label{alg:search_progressive}
\begin{algorithmic}[1] 
\STATE $\mathcal{X}\gets$ RandomInit$()$
\FORALL{$t\in[1,T]$}
    \STATE $i\gets \min(\lfloor mt/T\rfloor+1,m)$\label{line:reveal}
    \STATE $\mathcal{X}\gets$ Mutate$(\mathcal{X},\mathbbm{P}_f(y_1\dots y_i|x))$\label{line:mutate}
\ENDFOR
\RETURN $\mathcal{X}$
\end{algorithmic}
\end{algorithm}

More precisely, progressive search lets GA and PSO to evaluate any candidate input $x$ over a partial output $y_1\dots y_i$ (see Line \ref{line:mutate}). Since transformer-based language models compute the probability of the output by iterating over each symbols (see Equation \ref{eq:chain_rule}), generating only $i$ symbols reduces the computational cost. As the number of iterations $t$ increases, we generate more and more output symbols (see Line \ref{line:reveal}) until we recover the full objective function $\mathbbm{P}_f(y|x)$.

\subsection{Search Initialisation}
\label{sec:search_init}

As the search space for adversarial inputs is infinitely large, the choice of initialisation for both GA and PSO is crucial. Here, we focus on three main initialisation strategies:
\begin{itemize}
    \item \textbf{Random.} As a baseline, we experiment with random initialisation strategies for the population $\mathcal{X}$.
    \item \textbf{Output Copy.} Existing work on \textit{jailbreaks}, shows that it is sometimes possible to get a language models to repeat an input sequence~\cite{zou2023universal}. For this reason, we explore initialisation strategies that set $x\approx y$ for all elements $x\in\mathcal{X}$ of the population.
    \item \textbf{Inverted Model.} The work of \cite{Morris2023_language} trains an explicit inverted model $f^{-1}$ based on the T5 architecture. Accordingly, we initialise all $x\in\mathcal{X}$ by sampling $x=f^{-1}(y)$.
\end{itemize}
See Appendix \ref{sec:init_list} for more details.

\section{Preliminary Experiments}
\label{sec:experiments}

In this section, we present our empirical evidence. Here, we want to answer the following research questions:

\begin{itemize}
    \item \textbf{RQ1.} What is the most efficient search algorithm?
    \item \textbf{RQ2.} What is the impact of the initialisation strategy?
    \item \textbf{RQ3.} What is the relationship between weak and exact inversion?
\end{itemize}

\subsection{Experimental Setup}
\label{sec:exp_setup}

The code to replicate our experiments is available at: https://zenodo.org/doi/10.5281/zenodo.11069036 

\paragraph{Computational Infrastructure.} We use an NVIDIA's T4 16GB GPU for the experiments in Section \ref{sec:exp_init} and an NVIDIA's Quadro RTX 6000 24GB GPU for the rest.

\paragraph{Language Models.} We use the 124m parameter model GPT-2\footnote{huggingface.co/openai-community/gpt2} and the 7b parameter (quantized) model LLAMA-2-Chat\footnote{huggingface.co/TheBloke/Llama-2-7B-GGML}. For both, we set \texttt{temperature} to 0.7, \texttt{top\_p} to 0.95 and \texttt{top\_k} to 300.

\paragraph{Datasets.} We use a subset of Chatbot Arena Conversations\footnote{huggingface.co/datasets/lmsys/chatbot\_arena\_conversations}, which is part of the training set of the T5 inversion model in~\cite{Morris2023_language}. Specifically, we filter 30 input-output pairs from the dataset with the following features: the text is in English, the input is the first in the conversation, the input is under 15 tokens long, the output is under 100 tokens long, the target model has non-zero probability of generating the output and the output is not toxic according to the classifiers used. For the experiments in Section \ref{sec:exp_llm}, we remove the toxic filter and increase the input length to 64 tokens, resulting in a set of 50 input-output pairs.


\paragraph{Metrics.} For weak inversion, we measure the percentage of samples for which $\mathbbm{P}_f(x'|y) \geq \mathbbm{P}_f(x|y)$, where $x'$ is the best input found and $x$ is the original. For exact inversion, we measure both the percentage of strictly matching samples ($x'=x$), and the following fine-grained similarity metrics (borrowed from~\cite{Morris2023_language}): BLEU score \cite{papineni2002bleu}, token-level F1 score and cosine similarity according to the \texttt{text-embeddings-ada-02} model \cite{neelakantan2022text}. We repeat each experiment three times and report the mean and standard error.

\subsection{Search Algorithm Comparison}
\label{sec:exp_search}

\begin{table}[t] 
\centering
    \hspace{-2mm}
    \begin{tabular}{ llr|rr } 
    \toprule
    \multirow{2}{*}{Search} & \multirow{2}{*}{Objective} & \multirow{2}{*}{Obj. Calls} & \multicolumn{2}{c}{Weak Inversion}\\
    & & & Before & After \\
    \midrule
    \multirow{2}{*}{GA} & Full & 73K±0.2K & 13±3\% & 33±0\% \\ 
    & Progressive & 96K±0.2K & 18±5\% & 42±5\% \\ 
    \midrule
    \multirow{2}{*}{PSO} & Full & 42K±0.4K & 11±1\% & 27±0\% \\ 
    & Progressive & 46K±1.3K & 9±1\% & 31±2\% \\ 
    \bottomrule
    \end{tabular}
\caption{Weak inversion before and after searching for 350 minutes from random initialisation, objective function calls being an average.}
\label{tab:algo_table}
\end{table}

In Table \ref{tab:algo_table}, we compare text-based GA and embedding-based PSO search algorithms with both progressive and full objectives, under random initialisation. Given the vast difference in computational efficiency of these search algorithms, we terminate all of them after a given timeout of 350 minutes.

As expected, searching for adversarial examples improves the number of weak inversions. At the same time, text-based GA runs beats its embedding-based PSO counterpart by up to \~10 absolute points. Furthermore, switching to progressive search allows both GA and PSO to explore a larger portion of the search space, albeit with an approximate objective function, thus slightly improving their weak inversion capabilities.


\begin{tcolorbox}
\textbf{RQ1:} text-based GA with progressive objective is the most efficient search algorithm for weak inversion. 
\end{tcolorbox}


\subsection{Initialisation Comparison}
\label{sec:exp_init}

\begin{table*}[t]
\begin{adjustbox}{width=\textwidth, center}
    \begin{tabular}{ ll|rr|rrrrrrrr } 
    \toprule
    
    \multirow{2}{*}{Search} & \multirow{2}{*}{Initialisation} & \multicolumn{2}{c|}{Weak Inversion} & \multicolumn{2}{c}{Exact Inversion} & \multicolumn{2}{c}{BLEU} & \multicolumn{2}{c}{Token F1} & \multicolumn{2}{c}{Cos. Similarity}\\
    
    & & Before & After & Before & After & Before & After & Before & After & Before & After \\
    \midrule

    \multirow{6}{*}{GA} & Random & 13±3\:\:\% & 31±1\:\:\% & 0±0\% & 0±0\% & 0±0 & 0±0 & 0±0 & 1±0 & 70±0 & 72±0 \\
    
    
    
    
    & Output & \textbf{86±2\;\%} & \textbf{99±1\;\%} & 0±0\% & 0±0\% & 12±0 & 10±0 & 40±0 & 37±0 & 87±0 & \textbf{87±0} \\
    
    & Out. synonym & 46±17\% & 63±25\% & 0±0\% & 0±0\% & 4±0 & 3±0 & 30±0 & 28±0 & 87±0 & 86±0 \\
    
    & Out. paraphrase & 67±22\% & 73±18\% & 0±0\% & 0±0\% & 6±3 & 6±2 & 25±9 & 24±9 & 82±5 & 82±4 \\
    
    & Inversion & 53±0\:\:\% & 69±1\:\:\% & 0±0\% & 0±0\% & 15±0 & 14±0 & 38±0 & 35±0 & 85±0 & 84±0 \\
    
    & Inv. sample & 72±5\:\:\% & 83±4\:\:\% & 0±0\% & 0±0\% & \textbf{19±1} & \textbf{15±0} & \textbf{41±1} & \textbf{40±1} & \textbf{88±0} & \textbf{87±0} \\
    
    \midrule
    
    \multirow{6}{*}{PSO} & Random & 11±1\:\:\% & 27±2\:\:\% & 0±0\% & 0±0\% & 0±0 & 0±0 & 6±1 & 6±1 & 71±0 & 71±0\\
    
    
    
    
    & Output & 63±22\% & 63±22\% & 0±0\% & 0±0\% & 0±0 & 0±0 & \textbf{11±1} & 11±0 & 74±0 & \textbf{74±0}\\
    
    & Out. synonym & 63±3\:\:\% & 66±2\:\:\% & 0±0\% & 0±0\% & 0±0 & 0±0 & 8±1 & 9±0 & 73±0 & 72±0\\
    
    & Out. paraphrase & \textbf{83±0\;\%} & \textbf{83±0\;\%} & 0±0\% & 0±0\% & 0±0 & 0±0 & 11±0 & 10±0 & \textbf{74±0} & \textbf{74±0}\\
    
    & Inversion & 61±9\:\:\% & 64±8\:\:\% & 0±0\% & 0±0\% & 0±0 & 0±0 & 10±0 & 11±0 & 72±0 & 73±0\\
    
    & Inv. sample & 73±2\:\:\% & 73±2\:\:\% & 0±0\% & 0±0\% & 0±0 & 0±0 & 10±0 & \textbf{12±1} & 73±0 & 73±0 \\
    
    \bottomrule
    \end{tabular}
\end{adjustbox}

\caption{Inversion scores before and after searching for 200 minutes from different initialisations, using the full objective function.}
\label{tab:initial_table}
\end{table*}

In Table \ref{tab:initial_table}, we compare our search algorithms under a variety of different initialisation strategies. For further details on strategies and additional results, see Appendix \ref{sec:init_list} and \ref{sec:init_extra_results}. This set of experiments was run with a timeout of 200 minutes.

On the one hand, initialisation has a very large impact on the ability of GA and PSO to solve weak inversion. Interestingly, the most successful strategies involve copying the target output $y$ as the input $x$, either verbatim (\textit{Output}) or via some form of perturbation (\textit{Output synonym}, \textit{Output paraphrase}). Manual inspection of the generated inputs $x\in\mathcal{X}$ show that they retain most of the target output text $y$. Indeed, they differ only by the insertion of additional text around $y$. Though the additional text is often uninterpretable, we speculate that it is optimised to prompt the model to repeat the input, thus acting as a \textit{jailbreak}. 


On the other hand, we get the best exact inversion scores by using the explicit inverted model $f^{-1}$ from \cite{Morris2023_language} to initialise $\mathcal{X}$. In particular, the strategy of sampling many candidate inputs from $f^{-1}$ (\textit{Inversion sample}) seems to improve scores relative to when greedily sampling from $f^{-1}$ only once (\textit{Inversion}). At the same time, any amount of search, even by the best-performing text-based GA, makes the exact inversion metrics worse. Together, these two facts suggest that the weak and exact inversion objectives are indeed correlated, but not enough to act as surrogate objective functions. We comment further on this in Section \ref{sec:exp_llm}.



\begin{tcolorbox}
\textbf{RQ2:} initialisation has a larger impact on weak and exact inversion than the search algorithm. 
\end{tcolorbox}



\subsection{Language Model Comparison}
\label{sec:exp_llm}

\begin{figure}[t]
\centering
    \includegraphics[width=0.9\columnwidth]{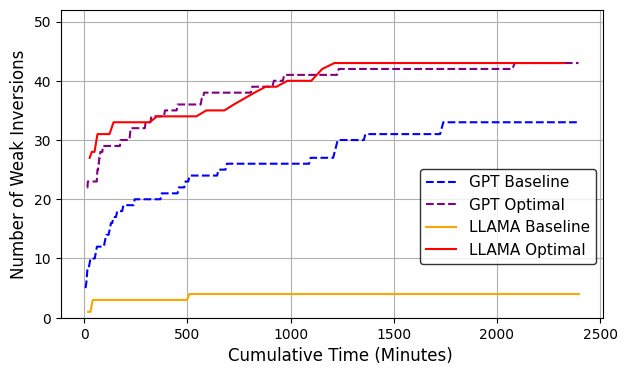}
\caption{Comparison between baseline and optimal GA search on different LLMs. The maximum possible weak inversion score is 50.}
\label{fig:weak_llm}
\end{figure}

In Figure \ref{fig:weak_llm}, we compare the effectiveness of our search on two different LLMs over a longer time frame. We test both a small (GPT-2) and a large (LLAMA-2-Chat) LLM. Furthermore, we show the performance gain of searching with the optimal hyper-parameters (text-based progressive GA with \textit{Inversion sample} initialisation) over the baseline parameters (GA with \textit{Random} initialisation and full objective).


In terms of weak inversion, we can see that the search continues to find improvements long after the timeouts of the experiments in Sections \ref{sec:exp_search} and \ref{sec:exp_init} (200 and 350 minutes, respectively), even though there are diminishing returns past 500 minutes. At the same time, the exact inversion scores (not shown in Figure \ref{fig:weak_llm}) remain zero in all settings even after searching for the whole 2400 minutes.

\begin{tcolorbox}
\textbf{RQ3:} weak inversion is not an effective surrogate objective for exact inversion.
\end{tcolorbox}

\section{Conclusions and Future Work}
\label{sec:conclusions}

In this paper, we show that searching for adversarial inputs for a specific target output does not improve our ability to reconstruct the original input that cause said output. Even though the two objectives of adversarial (weak) inversion and exact inversion seem to be mildly correlated, weak inversion cannot be used as a surrogate objective in a search algorithm. In the future, we plan to define a more effective surrogate objective, which might shed light on what the minimal amount of information that is required for exact inversion.





\bibliographystyle{named}
\bibliography{citations}

\section*{APPENDIX} 
\appendix

\begin{table*}[t] %
\begin{adjustbox}{width=\textwidth, center}
    \begin{tabular}{ ll|rr|rrrrrrrr } 
    \toprule
    
    \multirow{2}{*}{Search} & \multirow{2}{*}{Initialisation} & \multicolumn{2}{c|}{Weak Inversion} & \multicolumn{2}{c}{Exact Inversion} & \multicolumn{2}{c}{BLEU} & \multicolumn{2}{c}{Token F1} & \multicolumn{2}{c}{Cos. Similarity}\\
    
    & & Before & After & Before & After & Before & After & Before & After & Before & After \\
    \midrule
    
    \multirow{3}{*}{GA} & Rand. dataset &  16±1\:\:\% & 31±2\:\:\% & 0±0\% & 0±0\% & 0±0 & 0±0 & 4±1 & 3±0 & 70±0 & 73±0 \\
    
    & Rand. fluent &  17±0\:\:\% & 33±0\:\:\% & 0±0\% & 0±0\% & 0±0 & 0±0 & 6±0 & 4±0 & 71±0 & 72±0 \\
    
    & Rand. output &  58±1\:\:\% & 82±1\:\:\% & 0±0\% & 0±0\% & 5±0 & 4±0 & 24±1 & 25±0 & 83±1 & 83±0 \\

    \midrule
    
    \multirow{3}{*}{PSO} & Rand. dataset & 16±2\:\:\% & 27±0\:\:\% & 0±0\% & 0±0\% & 0±0 & 0±0 & 6±0 & 7±1 & 71±0 & 72±0\\
    
    & Rand. fluent & 23±7\:\:\% & 34±6\:\:\% & 0±0\% & 0±0\% & 0±0 & 0±0 & 8±1 & 8±1 & 71±0 & 71±0\\
    
    & Rand. output & 42±1\:\:\% & 44±1\:\:\% & 0±0\% & 0±0\% & 0±0 & 0±0 & 10±0 & 8±0 & 72±0 & 73±1\\
    
    \bottomrule
    \end{tabular}
\end{adjustbox}
\caption{Inversion scores before and after searching for 200 minutes from different initialisations, using the full objective function.}
\label{tab:initial_table_appendix}
\end{table*}

\section{GA and PSO Parameters} 
\label{sec:ga_pso_details}

We made use of the `eaSimple' implementation of the genetic/ evolutionary algorithm from the DEAP library, rather than the more specialised `eaMuPlusLambda' or `eaMuCommaLambda'. A single individual in the population would be represented by a variable-length list of numbers that range from 0 to the size of the target model's token vocabulary. We use the `cxUniform' implementation of mating strategy with independent probability set to 0.3. We use a custom implementation of mutation strategy with independent probability set to 0.1. In this case, 0.1 represents the probability of each token/ number in an individual's list receiving one mutation. The available mutations are changing the token value to a random value, inserting a random token to the left of the token, deleting the token or swapping the positions of this token with another in the text. Each mutation is equally likely to be chosen except for when only one token remains in the string (in which case you cannot delete nor swap). We use the `selTournament' implementation of the selection strategy with the explore-exploit variable of tournament size set to 15. The population size is set to 1000. \\

For particle swarm optimisation, individuals are represented as vectors of reals ranging from -1 to 1 and of 512 dimensions in size, both consequences of the embedding model we used. It was a T5 bottleneck autoencoder model\footnote{huggingface.co/thesephist/contra-bottleneck-t5-small-wikipedia} trained on english wikipedia articles. Its temperature was set to 1.0 and top\_p set to 0.9 during decoding, as recommended by the developer. The size  is fixed for the embeddings but not for the decoded output text, just like GA, though the sampling function does require a maximum sample length so we set this to 64 for the initial experiments and to 99 for secondary experiments (high above the original inputs maximum size in both cases). PSO has a predefined update function between iterations for which particle speeds are capped at a minimum of -0.5 and a maximum of 0.5. The Phi1 coefficient determines how much a particle's own best-known position influences its movement while the Phi2 coefficient determines how much the swarm's best-known position influences the particle's movement, both being set to 2.0 for balance in exploration-exploitation. The population size is set to 500.

\section{Initialisations}
\label{sec:init_list}

Descriptions of the strategies presented in Table \ref{tab:initial_table}:
\begin{itemize}
    \item \textbf{Random.} refers to randomly sampling from a uniform distribution to get a variably-long list of token IDs or a fixed-length embedding vector; 
    \item \textbf{Output.} refers to simply having the whole population start as the target output; 
    \item \textbf{Output synonym.} refers to starting with the target output after each word has been randomly replaced by one of its synonyms (which is likely to be different for each individual in the population) as provided by the WordNet corpus\footnote{wordnet.princeton.edu/documentation}; 
    \item \textbf{Output paraphrase.} refers to instead getting many variations of the target output by using a T5 model fine-tuned for paraphrasing\footnote{huggingface.co/Vamsi/T5\_Paraphrase\_Paws} - temperature being set to 1.5, top\_p being set to 0.99 and top\_k being set to 500; 
    \item \textbf{Inversion.} refers to giving the target output to the Morris et al. baseline inversion model\footnote{huggingface.co/wentingzhao/inversion\_seq2seq\_baseline} and using a single greedy sample from it for the whole population - temperature being set to 0.0; 
    \item \textbf{Inversion sample.} is similar to the previous, except you repeatedly sample from the model (something which the authors did not do themselves) to get variety among the population - temperature being set to 1.0, top\_p being set to 0.99 and top\_k being set to 500. 
\end{itemize}
Descriptions of the strategies presented in Table \ref{tab:initial_table_appendix}:
\begin{itemize}
    \item \textbf{Random dataset.} refers to randomly sampling from an out-of-distribution dataset, specifically a collection of tweets made in February 2024\footnote{kaggle.com/datasets/fastcurious/twitter-dataset-february-2024}; 
    \item \textbf{Random fluent.} refers to randomly choosing a single token and then using the target model to generate the rest of each input; 
    \item \textbf{Random output.} is similar but you instead start with the target output sequence to encourage the following text to be of a similar theme; 
\end{itemize}
Note that PSO requires an additional step of converting the described initialisation text to an embedding.

\section{Additional Results}
\label{sec:init_extra_results}

\begin{figure*}[t]
\centering
    \begin{subfigure}[t]{0.48\textwidth}
        \centering
            \includegraphics[height=4.4cm]{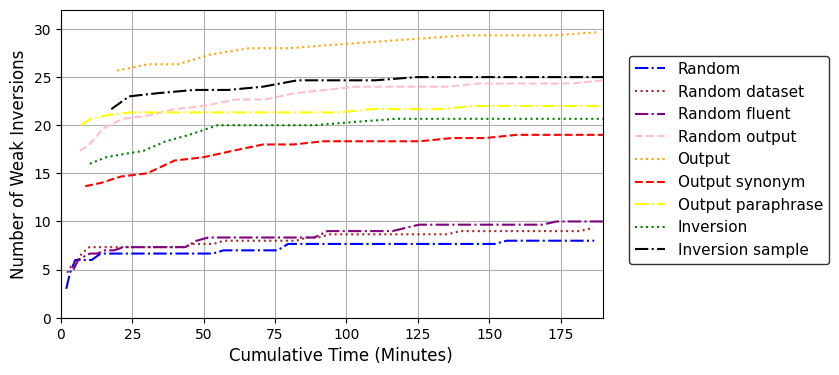}
        \caption{Text-based GA.}
        \label{fig:text_initial_appendix}
    \end{subfigure}
    \hfill
    \begin{subfigure}[t]{0.48\textwidth}
        \centering
            \includegraphics[height=4.4cm]{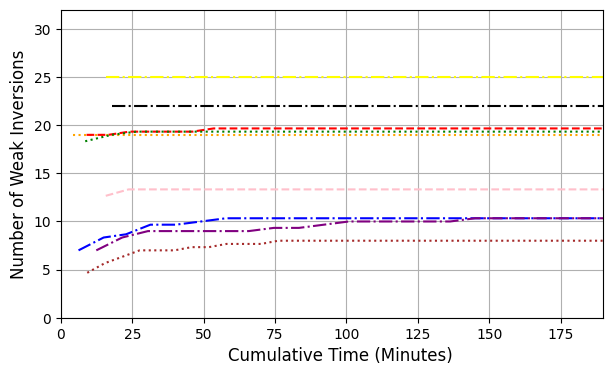}
        \caption{Embedding-based PSO.}
        \label{fig:embedding_initial_appendix}
    \end{subfigure}
\caption{Inversion scores for different search algorithms and initialisations, using the full objective function.}
\label{fig:initial_appendix}
\end{figure*}

In Table \ref{tab:initial_table_appendix}, we compare our search algorithms under a few additional initialisation strategies (also detailed in Appendix \ref{sec:init_list}). Here, we explore how important it is for input text to be sampled from a distribution of syntactically-correct English, which is separate from it's semantic relevance to the target input. Text from \textit{Random dataset} is `correct' in terms of it being accepted by English readers while text from \textit{Random fluent} is `correct' in terms of it being accepted by our own LLM (i.e. it producing the text itself means that the text has a low perplexity score). Both of these perform slightly better than \textit{Random} but not by much and they are equivalent in terms of input similarity metrics. However, there is a significant improvement over \textit{Random} for \textit{Random output}, which reaffirms previous conclusions that relevance to the target's semantics is much more important than other factors. The difference is not as significant for PSO as it is for GA, but this is also in line with previous results which showed that PSO can at most produce a few percent gain in weak inversion scores for initialisations that are not random (i.e. the worst performing ones). Either way, \textit{Random output} scores still do not beat the simple \textit{Output} initialisation, showing that the two are meaningfully different. \\

In Figure \ref{fig:text_initial_appendix} and \ref{fig:embedding_initial_appendix}, we present the broader picture of weak inversion scores progressing over cumulative time for all initialisation runs. Note that each line represents the mean value across each run, with error bars excluded for visual clarity. Something which could not be seen before is that the lines begin at differing times. This captures the amount of processing required to generate each initialisation as well as the time it takes to do an initial evaluation of each individual in the population, the latter being dependent on the length of the text being evaluated as well as whether the evaluation stops early due to some target token having a zero probability of being output. This is why we find that the simple and badly-performing \textit{Random} lines begin the soonest, while the simple but well-performing \textit{Output} lines begin later. The latter result is not as clear for PSO, which requires an additional encoding-decoding step at initialisation and for which \textit{Output} scores lower. Tangentially, runs for which evaluation processes are faster are also also able to get through more iterations of optimisation. although values continue to increase for GA, we do see that all gradients declines over time, as in Figure \ref{fig:weak_llm}. This is especially clear for PSO, though the majority of its lines are entirely flat due to its ineffectiveness to improve on the initialisation. Notably, weak inversion scores local to each generation's population are much more sporadic than the 'best so far' scores presented here.



\end{document}